# Phase diagram, ferromagnetic martensitic transformation and magnetoresponsive properties of Fe-doped MnCoGe alloys

G. J. Li[a], E. K. Liu[a], H.G. Zhang[a], Y. J. Zhang[a], J. L. Chen[a], W. H. Wang[a], H. W. Zhang[a], G. H. Wu[a,*], S. Y. Yu[b]

[a] Beijing National Laboratory for Condensed Matter Physics, Institute of Physics, Chinese Academy of Science, Beijing 100190, People's Republic of China

[b] School of Physics, Shandong University, Jinan 250100, People's Republic of China

Abstract

The crystal structure and magnetoresponsive properties of Fe-doped MnCoGe alloys have been investigated using x-ray diffraction (XRD) and magnetic measurements. By alloying the Fe-containing isostructure compounds into MnCoGe, a magnetostructural transition from paramagnetic austenite to ferromagnetic martensite with large magnetization difference can be realized in a temperature window between the Curie temperatures of the austenite and martensite, resulting in magnetic-field-induced martensitic transformations and large magnetic-entropy changes. A structural and magnetic phase diagram of Fe-doped MnCoGe alloys has been proposed.

1.Introduction

Ferromagnetic (FM) martensitic-transition materials have attracted increasing attention due to their magnetoresponsive effects, including magnetic-field-induced martensitic transformation/strain effect[1-3], magnetoresistance[4, 5] and magnetocaloric effect[6, 7], et al. For these properties, the magnetostructural coupling between structural and magnetic transition plays a crucial role. Thus, exploring materials with magnetostructural transition is of importance for fundamental science and technologic application.

The magnetic equiatomic MM'X (M, M'=transition metals, X=Si, Ge, Sn) compounds [8-11] have become interesting research objects due to their remarkable magnetoresponsive properties[12-25]. As one of the important MM'X compounds, the stoichiometric MnCoGe alloy transforms from the Ni$_2$In-type hexagonal structure (space group p6$_3$/mmc, 194) to the TiNiSi-type orthorhombic structure (space group pnma,62) via the martensitic transformation at about 420 K, i.e. martensitic transformation temperature ($T_m$). The Curie temperatures of hexagonal austenite and orthorhombic martensite are 276 K ($T_C^A$) and 355 K ($T_C^M$), respectively [9-11]. It can be seen that $T_C^M$ is about 80 K higher than $T_C^A$. Because $T_m$ is higher than $T_C^M$, the martensitic transformation occurs in the paramagnetic (PM) state without a magnetostructural coupling. If lowering the $T_m$ below $T_C^M$, a magnetostructural transition can occur in the magnetic region. In previous studies, a martensitic transformation between PM Ni$_2$In-type structure and FM TiNiSi-type structure can be obtained in a temperature window between $T_C^A$ and $T_C^M$ by introducing the vacancy[17], the fourth elements[18, 19, 21, 22, 24, 25] or pressure[26-28] in MnCo(Ni)Ge alloys. Thus, a strong magnetostructural coupling with large magnetization difference can be realized. The previous reports indicate that the doping of Fe in Heusler alloys[29-31], Ru-based[32] alloys or Gd$_5$Ge$_2$Si$_2$[33] can suppress the martensitic transition and stabilize the parent phases. Recently, chemically alloying the Fe-containing isostructure MnFeGe or FeNiGe alloy to Ni$_2$In-type MnNiGe is proved to be an effective method to manipulate the first-order martensitic transformation in a temperature window [18]. Another study has reported the magnetocaloric effect based on a second-order magnetic transition at $T_C$ in MnCo$_{1-x}$Fe$_x$Ge alloys (x=0-1.00)[34]. However, the study on the first-order martensitic transformation based on the coupling of magnetic and structural transitions, and their magnetoresponsive properties of MnCo$_{1-x}$Fe$_x$Ge($x$<0.10) is still absent up to now.

In this work, based on the isostructural alloying method, the MnCo$_{1-x}$Fe$_x$Ge and Mn$_{1-x}$Fe$_x$CoGe alloys were prepared. The study on the first-order FM martensitic transformation in Fe-doped MnCoGe alloys was carried out. By Fe substitution, a temperature window between $T_C^A$ and $T_C^M$ and therein a martensitic transformation from PM austenite to FM martensite with large magnetization difference were realized. The magnetic-field-induced martensitic transformations and magnetocaloric effects were also investigated.

2. Experimental method

Polycrystalline ingots of MnCo$_{1-x}$Fe$_x$Ge($x$ = 0-0.15) and Mn$_{1-x}$Fe$_x$CoGe($x$ = 0-0.10) alloys were prepared by arc-melting high-purity metal Mn, Co, Ge and Fe with purity of 99.99% and higher under argon atmosphere. The ingots were homogenized subsequently in evacuated quartz tubes with argon at 1123 K for five days and then quenched into ice water. The X-ray diffraction (XRD) with Cu-Kα was employed to characterize the crystal structures. The magnetic measurements of the samples were carried out by a superconducting quantum interference device (SQUID).

3. Results and discussion

Figure 1 (a) and (b) show the powder XRD patterns of MnCo$_{1-x}$Fe$_x$Ge and Mn$_{1-x}$Fe$_x$CoGe alloys measured at room temperature, respectively. It can be seen that the crystal structure changes from the coexistence of TiNiSi-type and Ni$_2$In-type structures to the single Ni$_2$In-type structure with the substitution of Fe for Co or Mn atoms. This indicates the martensitic transformation temperatures of the two series of alloys decrease from higher temperature down to lower temperature. The substitution of Fe for Mn or Co can effectively change the phase stability of MnCoGe alloys.

Indexing the XRD pattern of the Mn$_{0.97}$Fe$_{0.03}$CoGe alloy (not shown in Fig.1(b)), the lattice parameters of Ni$_2$In-type hexagonal austenitic and TiNiSi-type orthorhombic martensitic phase were obtained. For austenitic structure, the lattice parameters are a =4.0812 Å, c=5.289 Å, and those of martensitic structure are a=5.9074 Å, b=3.8178 Å, c=7.0445 Å, respectively. The unit-cell volume increases by 4.2% during the martensitic transformation, indicating a large lattice distortion. This large and positive volume expansion can result in significant difference of atomic surrounding and crystal structure between $Ni_2In$-type and TiNiSi-type structures.

It is known that the radius of Fe (0.172 nm) is larger than that of Co atom (0.167 nm) while smaller than that of Mn atom (0.179 nm)[35]. In $Ni_2In$-type structure, however, the main peaks of two series of alloys both shift to the large-angle direction with the substitution of Fe for Mn or Co atoms, which means the unit-cell sizes of the two series of alloys all decrease with the increasing of Fe content. In $MnCo_{1-x}Fe_xGe$ alloys, the newly introduced Fe atoms will occupy Co sites since both MnCoGe and MnFeGe are isostructural[36]. Thus the strong covalent effect between Fe and Ge atoms formed. The reduction of lattice parameter can be ascribed to this enhanced covalent effect, which just like the case in Fe-doped MnNiGe alloys[18].

The temperature dependence of magnetization for $MnCo_{1-x}Fe_xGe$ and $Mn_{1-x}Fe_xCoGe$ alloys measured under a magnetic field of 100 Oe were shown in Fig.2 (a) and (b), respectively. It can be seen that the martensitic transformation temperature decreases with the substitution of Fe for Co or Mn atoms, which means the doping of Fe atoms can realize the desired coupling of martensitic structural transition and magnetic transition. This effect is very similar to the case in isostructural Fe-doped MnNiGe alloys [18], in which the doping of Fe for transition-metal atoms can also effectively decrease the martensitic transition temperature to magnetic ordering temperature. In $Mn_{1-x}V_xCoGe$[21] and $MnNi_{1-x}Co_xGe_{1.05}$[23] alloys, the researchers attributed this effect to the so-called "chemical pressure", originating from the different atom radii during substitution. The "chemical pressure" works like the applied external hydrostatic pressure to tune the phase stability[26]. In this work, the substitution of Fe (0.172 nm) for Co (0.167 nm) or Mn (0.179 nm) atoms can give rise to the local lattice distortion in $MnCo_{1-x}Fe_xGe$ and $Mn_{1-x}Fe_xCoGe$ alloys, consequently imposing a similar "chemical pressure" effect and modifying the relative stability between the austenitic and martensitic phases. In nature, the "chemical pressure" corresponds to the local chemical bonding in these alloys. In Fe-doped MnCoGe alloys, as demonstrated in Fe-doped MnNiGe alloys[18], the strengthened covalent bonding between the nearest-neighbor Fe and Ge atoms with respect to the nearest-neighbor Mn/Co and Ge atoms increases the stability of $Ni_2In$-type austenite structure and consequently decreases the martensitic transformation temperature.

The thermomagnetization curves of the $MnCo_{0.94}Fe_{0.06}Ge$ and $Mn_{0.97}Fe_{0.03}CoGe$ alloys in a magnetic field up to 50 KOe are shown in the insets of Fig.2 (a) and (b), respectively. Due to the PM/FM-type martensitic transformation, the magnetization difference values for the two alloys are about 51 and 58 emu/g, respectively, which is closed to that of $Mn_{1-x}CoGe$[17]. This magnetization difference may give rise to the magnetic-field-induced martensitic transformation and the large magnetic-entropy changes in the two series of alloys around room temperature.

Meanwhile, comparing the thermomagnetization curves of $MnCo_{0.94}Fe_{0.06}Ge$ with $Mn_{0.97}Fe_{0.03}CoGe$ alloys under 100 Oe and 50 KOe, we can find that their martensitic starting transition temperatures were pushed upward about 10 K by a magnetic field of 50 kOe, which implies a magnetic-field-induced martensitic transformation occurs. This also suggests that the stability of martensitic phase can be enhanced with the aid of applied external magnetic field.

Collecting the magnetic measurement data, the structural and magnetic phase diagrams for $MnCo_{1-x}Fe_xGe$ and $Mn_{1-x}Fe_xCoGe$ alloys are both shown in Fig. 3. By alloying isostructural MnFeGe or FeCoGe compounds into MnCoGe, two interesting results can be observed as follows: (i) In the present change range of composition, the Curie temperatures of austenite phase ($T_C^A$) and martensite phase ($T_C^M$) remain almost constant and they are basically at 276 K and 355 K, respectively, which are consistent with those of MnCoGe alloy[9, 11, 17]. (ii) The martensitic transformation temperature is lowered, spanning over $T_C^A$ and $T_C^M$. These two achievements subsequently result in a temperature window with an interval width of about 80 K between $T_C^A$ and $T_C^M$. In this temperature window, the martensitic transformations with large magnetization difference can occur in $Mn_{1-x}Fe_xCoGe$ ($x \leq 0.05$) and $MnCo_{1-x}Fe_xGe$ ($x \leq 0.06$) alloys.

In what follows, the magnetoresponsive properties of $MnCo_{1-x}Fe_xGe$ and $Mn_{1-x}Fe_xCoGe$ alloys are presented. The isothermal magnetization (*M-H*) curves of $MnCo_{0.94}Fe_{0.06}Ge$ and $Mn_{0.97}Fe_{0.03}CoGe$ alloys were measured in the process of increasing and decreasing magnetic field up to 50 KOe across the temperature windows. These results are shown in Fig.4 (a) and (b), respectively. For both alloys, their magnetizations show similar changes under the applied magnetic field in the temperature windows. Above the $T_m$, the magnetization linearly increases with the increase of magnetic field, which indicates the PM ground state of $Ni_2In$-type structure. While below the $T_m$, the two alloys with TiNiSi-type structure show typical FM behavior. Around the $T_m$, an S-shaped metamagnetic transition with a hysteresis indicates the occurrence of the remarkable magnetic-field-induced martensitic transformation from PM austenite to FM martensite. This is due to the larger Zeeman energy in the FM martensite introduced by the applied external magnetic field[3, 4, 17].

Based on the isothermal magnetization curves, the magnetic-entropy changes for $MnCo_{0.94}Fe_{0.06}Ge$ and $Mn_{0.97}Fe_{0.03}CoGe$ alloys were estimated by Maxwell relation[37], as shown below, and shown in Fig.5 (a) and (b), respectively.

$$\Delta S_m(T,H) = \int_0^H \left(\frac{\partial M(T,H)}{\partial T}\right)_H dH \tag{1}$$

For $MnCo_{0.94}Fe_{0.06}Ge$ and $Mn_{0.97}Fe_{0.03}CoGe$ alloys, the largest magnetic-entropy changes ($\Delta S_m$) are about -27.5 and -10.6 J/Kg·K under the applied magnetic field of 50 KOe around the martensitic transformation temperature, respectively. Apparently, compared with $MnCo_{0.94}Fe_{0.06}Ge$ alloy, the $Mn_{0.97}Fe_{0.03}CoGe$ alloy shows a relatively smaller magnetic-entropy change in a wide temperature range. And the value of magnetic-entropy change of $MnCo_{0.94}Fe_{0.06}Ge$ alloy can be comparable with the those of $Mn_{1-x}CoGe$ (-26 J/Kg·K for $\Delta H=50$ kOe)[17], $Ni_{52.6}Mn_{23.1}Ga_{24.3}$ (-18.0 J/Kg·K for $\Delta H=50$ kOe)[38] and $Ni_{50}Mn_{33.13}In_{13.90}$ (28.6 J/Kg·K for $\Delta H=50$ kOe)[39]. These negative $\Delta S_m$ values in this work can be ascribe to the first-order ferromagnetic martensitic transformation from high-temperature PM austenite to the low-temperature FM martensite. Importantly, just like the case in Fe-doped MnNiGe alloys, these martensites are in a more ferromagnetic ordered state after the phase transition, which results in a large $\Delta S_m$ value in Fe-doped MnCoGe alloys. Thus, for $MnCo_{0.94}Fe_{0.06}Ge$ and $Mn_{0.97}Fe_{0.03}CoGe$ alloys, the large and negative $\Delta S_m$ values around room temperature enables them to be potential magnetoresponsive materials. For the difference of magnetic-entropy changes between $MnCo_{0.94}Fe_{0.06}Ge$ and $Mn_{0.97}Fe_{0.03}CoGe$, it can be ascribed to their different magnetization behavior under the same magnetic field. As shown in Fig.2(a) and (b), the martensitic transformation of $MnCo_{1-x}Fe_xGe$ in a narrower temperature range is sharper than that of $Mn_{1-x}Fe_xCoGe$ alloys, which corresponds to higher value of $\left(\frac{\partial M(T,H)}{\partial T}\right)_H$. Thus, based on Maxwell relation, the $\Delta S_m$ of $MnCo_{1-x}Fe_xGe$ alloys is higher than those of $Mn_{1-x}Fe_xCoGe$ alloys.

4. Conclusion

The crystal structure, ferromagnetic martensitic transformation and magnetoresponsive properties of $MnCo_{1-x}Fe_xGe$ and $Mn_{1-x}Fe_xCoGe$ alloys have been investigated in this work. Using isostructural alloying method, the martensitic transformation from $Ni_2In$-type structure to TiNiSi-type structure can be lowered down to the magnetic ordering temperatures of the two series of alloys. A Curie-temperature window can be uncovered by substituting of Fe for Co or Mn.

The desired PM/FM-type magnetostructural transition, accompanied with a large magnetization difference, can be thus realized in this temperature window between $T_C^M$ and $T_C^A$. The remarkable magnetic-field-induced martensitic transformation and the large and negative magnetic-entropy changes around room temperature were observed in $MnCo_{0.94}Fe_{0.06}Ge$ and $Mn_{0.97}Fe_{0.03}CoGe$ alloys.

ACKNOWLEDGMENT

This work is supported by the National Natural Science Foundation of China (Grant Nos. 11174352 and 50901043) and National Basic Research Program of China (973 Program: 2012CB619405).
.

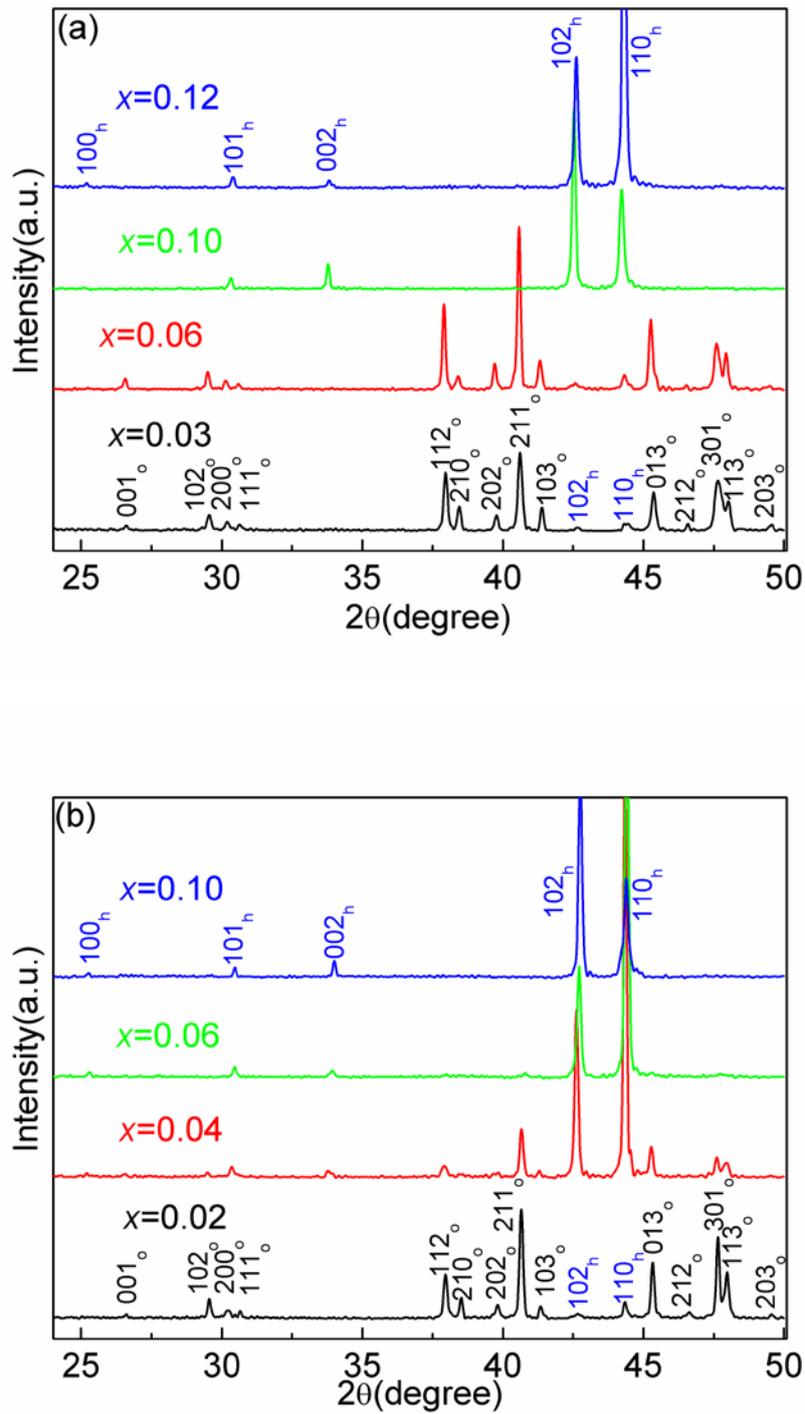

Fig.1. (Color online) XRD patterns of MnCo$_{1-x}$Fe$_x$Ge alloys ($x$ = 0.03, 0.06, 0.10, 0.12) (a) and Mn$_{1-x}$Fe$_x$CoGe alloys ($x$ = 0.02, 0.04, 0.06, 0.10) (b). Here h and o denote the Ni$_2$In-type hexagonal and TiNiSi-type orthorhombic structures, respectively.

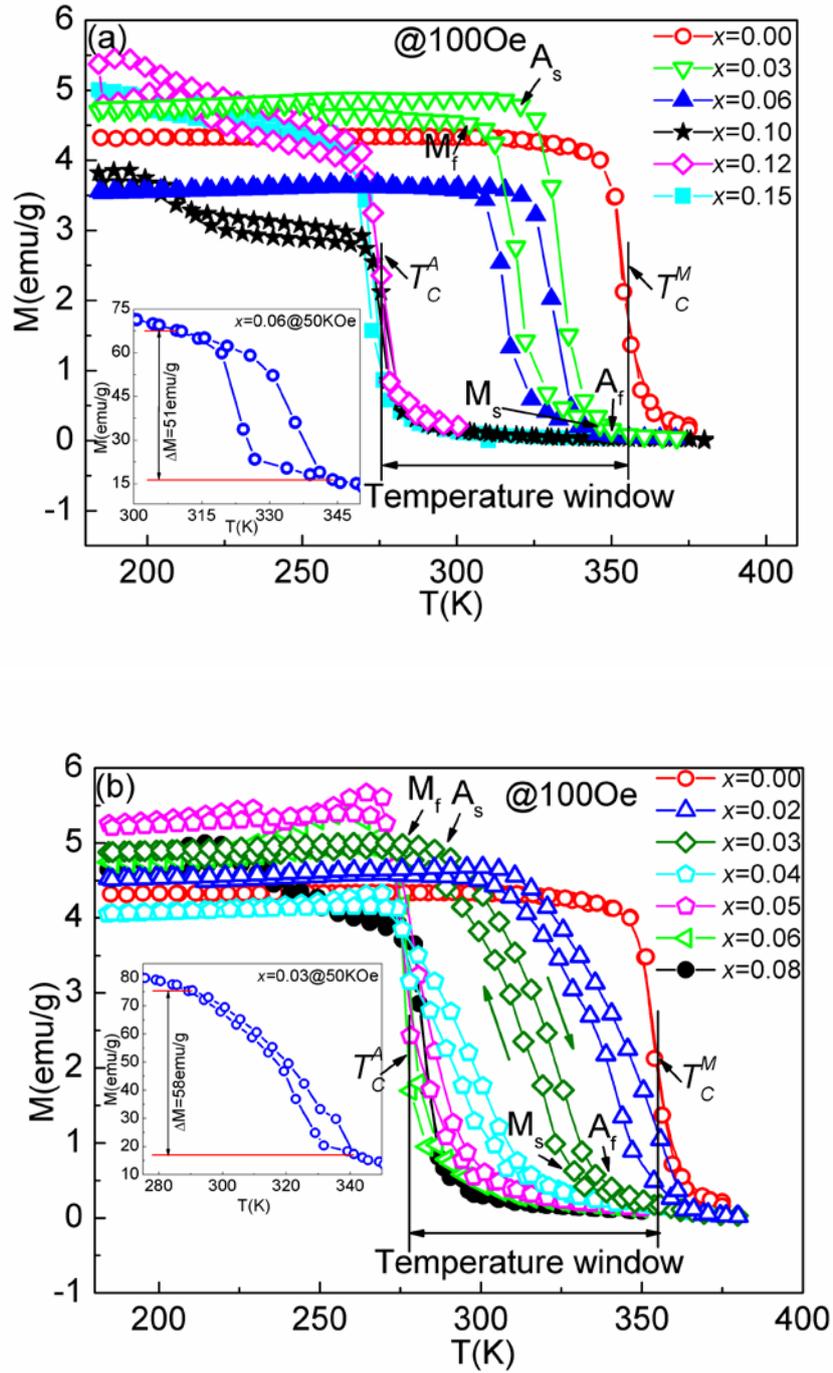

Fig.2. (Color online) Temperature dependence of magnetization of MnCo$_{1-x}$Fe$_x$Ge alloys (a) and Mn$_{1-x}$Fe$_x$CoGe alloys (b) measured under a magnetic field of 100 Oe. Insets of (a) and (b) show the temperature dependence of magnetization of MnCo$_{0.94}$Fe$_{0.06}$Ge and Mn$_{0.97}$Fe$_{0.03}$CoGe alloys under a magnetic field of 50 kOe, respectively. $M_s$ and $M_f$ denote the starting and finishing temperatures of the martensitic transformation; $A_s$ and $A_f$ are the starting and finishing temperatures of the reverse transformation.

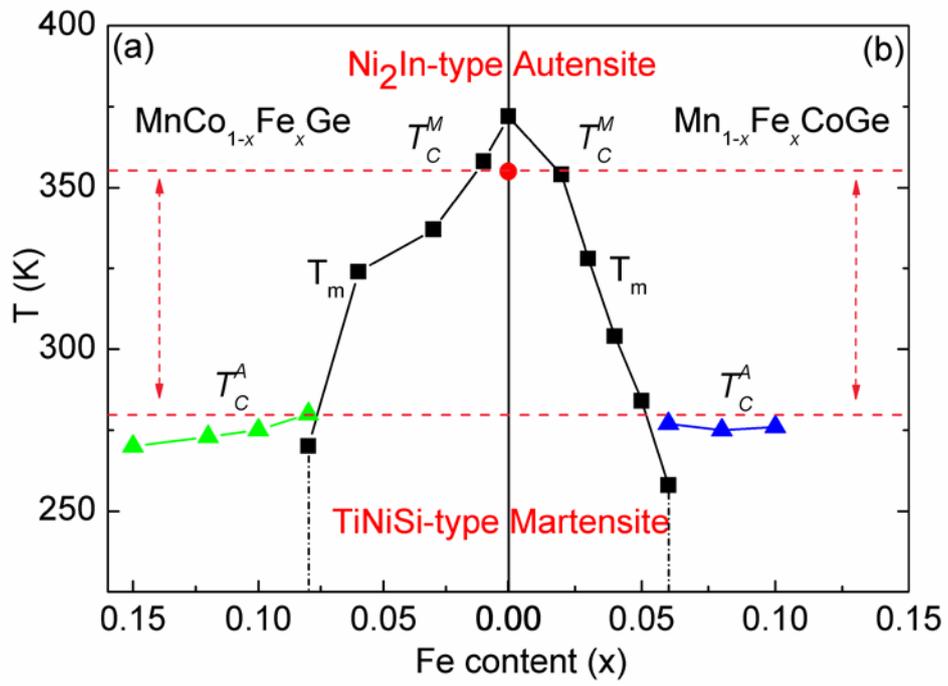

Fig.3. (Color online) Structural and magnetic phase diagram of MnCo$_{1-x}$Fe$_x$Ge (a) and Mn$_{1-x}$Fe$_x$CoGe alloys (b).

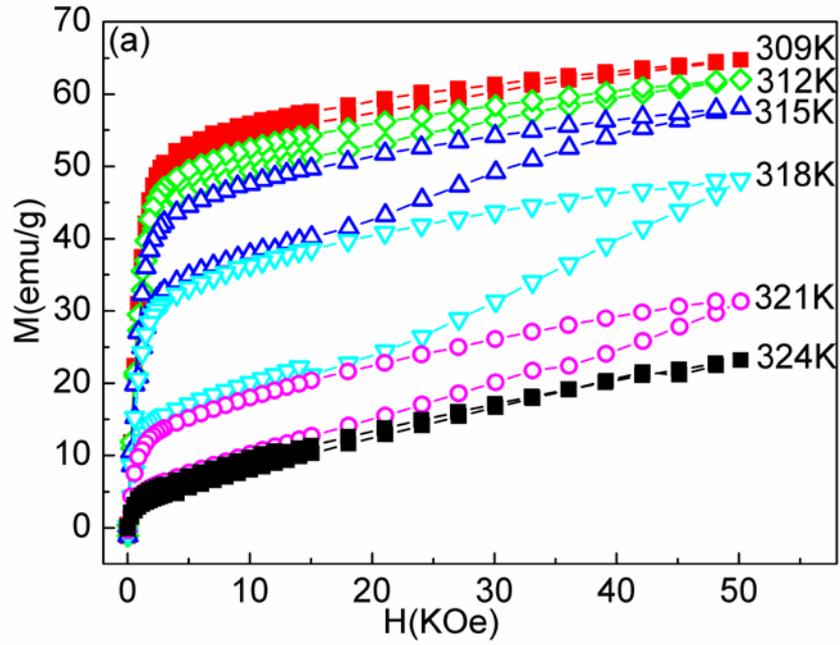

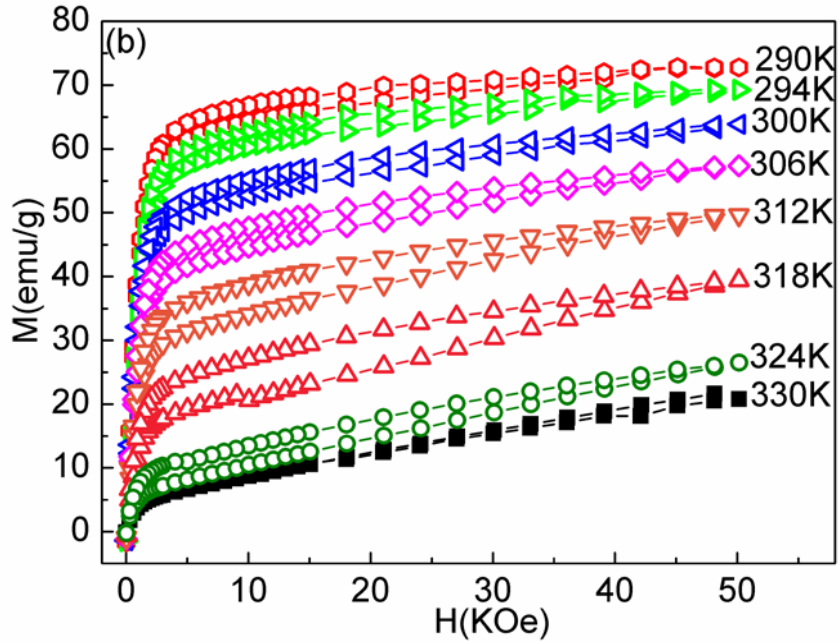

Fig.4. (Color online) Isothermal magnetization curves of the MnCo$_{0.94}$Fe$_{0.06}$Ge (a) and Mn$_{0.97}$Fe$_{0.03}$CoGe (b) alloys at various temperatures across the martensitic transformation in a magnetic field up to 50 kOe.

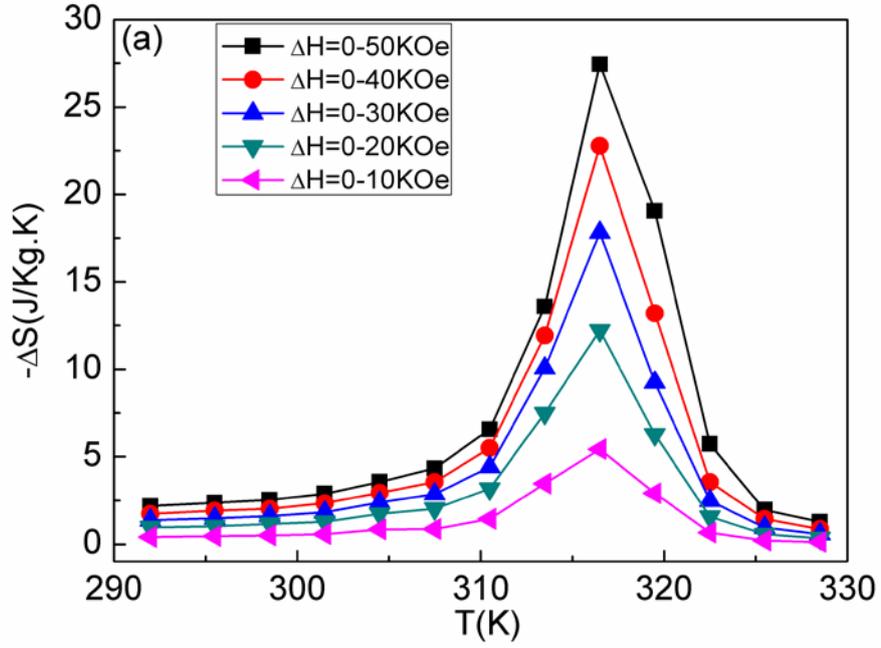

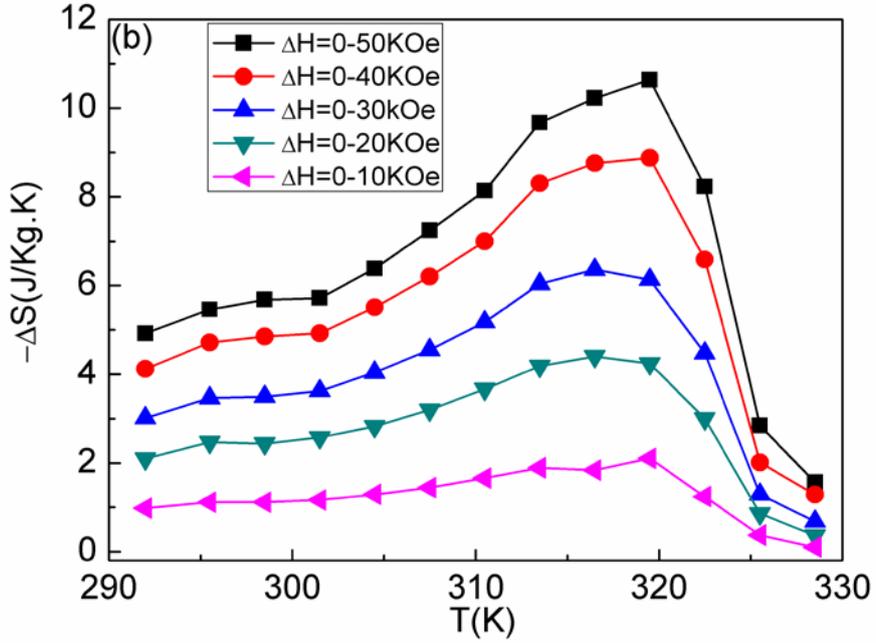

Fig.5. (Color online) Isothermal magnetic-entropy changes derived from isothermal magnetization curves of MnCo$_{0.94}$Fe$_{0.06}$Ge (a) and Mn$_{0.97}$Fe$_{0.03}$CoGe alloys (b).